# On the origin of the above-room-temperature magnetism in the 2D van der Waals ferromagnet Fe$_3$GaTe$_2$

Authors: Alberto M. Ruiz[1], Dorye L. Esteras[1], Diego López-Alcalá[1] and José J. Baldoví[1,*]

[1]Instituto de Ciencia Molecular, Universitat de València, Catedrático José Beltrán 2, 46980 Paterna, Spain. E-mail: j.jaime.baldovi@uv.es

**ABSTRACT**

Recent advancements in 2D magnetic materials have attracted a growing interest driven by their unique properties and potential applications in spintronic devices. However, the scarcity of systems that exhibit magnetism at room-temperature has limited their practical implementation into functional devices. In this work we focus on the recently synthetised van der Waals (vdW) ferromagnet Fe$_3$GaTe$_2$, which exhibits above-room-temperature magnetism (T$_c$ = 350-380 K) and strong perpendicular magnetic anisotropy. Through first-principles calculations, we examine the magnetic properties of Fe$_3$GaTe$_2$ and compare them with the widely known Fe$_3$GeTe$_2$ ferromagnet. Our calculations unveil the complex microscopic mechanisms governing their magnetic behaviour, emphasizing the pivotal role of the ferromagnetic in-plane exchange interactions in the stabilization of the elevated T$_c$ in Fe$_3$GaTe$_2$. Additionally, we predict the stability, strong perpendicular anisotropy and high T$_c$ of single-layer Fe$_3$GaTe$_2$. We also demonstrate the potential of strain engineering and electrostatic doping to modulate its magnetic exchange interactions and anisotropy. Our results incentivise the isolation of the monolayer and pave the way for the future optimization of Fe$_3$GaTe$_2$ in magnetic and spintronic nanodevices.

**INTRODUCTION**

The recent breakthrough of long-range magnetic order in two-dimensional (2D) van der Waals (vdW) magnetic materials represents a particularly exciting research avenue that opens new pathways for exploring physical phenomena and emerging applications. [1–3] The discovery of ferromagnetic materials at the 2D limit, such as the semiconductors $CrI_3$ and $Cr_2GeTe_6$, has attracted a lot of attention since 2017 [1,2]. However, practical applications are drastically hindered by their low critical temperatures below liquid nitrogen temperature [4,5]. In this regard, materials such as CrSBr or the metallic $Fe_3GeTe_2$ have come to the forefront, retaining magnetic properties down to monolayer thickness with higher Curie temperatures ($T_C$) of 146 and 130 K, respectively [6–8]. In this context, external fields, strain engineering and electrostatic doping have proven to be powerful strategies to improve the magnetic behaviour of these systems [4,9–13].

In particular, $Fe_3GeTe_2$ is a metal distinguished by its resilience after exfoliation [14] and exhibits a high $T_C$ with substantial strong magnetic anisotropy [15,16], suggesting its potential for spintronic applications [17–20]. While bulk-phase magnetism persists up to 230 K, $T_C$ is significantly reduced in thinner layers [6,14]. To address this, several strategies have been employed, such as electron gating [14], intercalation [21–23], or external pressure [24], to name a few. Among them, increasing the Fe composition in $Fe_xGeTe_2$ crystals has proven to be a powerful method to significantly boost the $T_c$, reaching up to 310 K [25–27].

More recently, bulk $Fe_3GaTe_2$ has emerged as a magnetic material of interest due to its above-room-temperature magnetism (350-380 K) and strong perpendicular magnetic anisotropy [28]. Its electronic structure has been deeply investigated [29], it has been incorporated into spin-valve devices [30–32] and has shown potential for hosting magnetic skyrmions at room-temperature [33–35]. Nevertheless, a comprehensive magnetic analysis to elucidate the main mechanisms governing the high critical temperature of $Fe_3GaTe_2$ and the possibility of maintaining its outstanding magnetic properties at the monolayer limit are still lacking.

In this work, through first-principles calculations combined with a Wannier-based tight-binding model, we investigate the fundamental mechanism responsible for the room-temperature magnetism in bulk $Fe_3GaTe_2$. Furthermore, we present a comparative analysis with $Fe_3GeTe_2$, unambiguously demonstrating the pivotal role of the competing ferromagnetic-antiferromagnetic in-plane couplings on the critical temperatures of both systems. In addition, we prove the dynamical stability of $Fe_3GaTe_2$ monolayer and explore the impact of strain engineering and electrostatic doping in the modification of the magnetic and electronic properties of this system.

## RESULTS AND DISCUSSION

Fe$_3$GaTe$_2$ is a van der Waals (vdW) magnetic material, isostructural to the widely known Fe$_3$GeTe$_2$, that crystallizes within a hexagonal layered structure in the space group P6$_3$/mmc (N$_0$. 194). Each unit cell contains two layers of the material stacked along the *c* direction (with an *AB* stacking pattern), which are separated by an interlayer spacing of ~7.8 Å. The lattice parameters are a = b = 3.986 Å and c = 16.229 Å [28]. Within each layer the system is formed by five atomic sublayers containing two external surfaces of Te atoms, a pair of internal layers formed by two equivalent Fe atoms (Fe$_1$ and Fe$_2$) and a central layer consisting in Ga and inequivalent Fe$_3$ atoms (Figure 1(a)).

Firstly, we perform first-principles calculations on bulk Fe$_3$GaTe$_2$. The results reveal magnetic moments of 2.05 and 1.35 $\mu_B$/Fe for the equivalent Fe$_{1,2}$ and the inequivalent Fe$_3$ atoms, respectively. Ga and Te showed negligible polarization. The calculated average magnetic moment is 1.82 $\mu_B$/Fe atom, consistent with experimental observations at 3 K and previous theoretical studies [28,36]. Furthermore, we extract the magnetic anisotropy energy (MAE) by means of the energy difference between in-plane and out-of-plane magnetic configurations (where a positive value indicates an out-of-plane magnetization easy axis). Our results yield a value of MAE of 0.31 meV/Fe, being in good agreement with previous studies [29,36] and confirming that the material exhibits a strong perpendicular anisotropy with spins aligned along the *c* axis.

To explore the reasons behind the different T$_c$ of Fe$_3$GaTe$_2$ and Fe$_3$GeTe$_2$, which are 380 and 240 K, respectively, we determine the magnetic exchange couplings in both materials and perform atomistic simulations (see Computational details). The exchange interactions are determined by first formulating a tight-binding Hamiltonian in the basis of maximally localized Wannier Functions (MLWFs) [37], and then employing the TB2J package [38], which is based in the use of Green's functions. The spin Hamiltonian has the following form:

$$H = -\sum_{i \neq j} J_{ij} \vec{S}_i \cdot \vec{S}_j - \sum_i A(\vec{S}_i^z)^2$$

where J$_{ij}$ represent the isotropic exchange interactions, S$_i$ and S$_j$ are the magnetic moments of the different sites and A accounts for the magnetic anisotropy of the system.

Our calculations reveal that the magnetic interaction picture of Fe$_3$GaTe$_2$ and Fe$_3$GeTe$_2$ is quite intricate, where the emergent global ferromagnetism arises from a competition between different couplings. We categorize these interactions into two groups, namely (*i*) inter-plane interactions, which encompass J$_{12}$

(interactions between atoms Fe$_1$-Fe$_2$) and J$_{13}$ (Fe$_1$-Fe$_3$), as well as (*ii*) in-plane exchanges, comprising J$_{11}$ (Fe$_1$-Fe$_1$) and J$_{33}$ (Fe$_3$-Fe$_3$) (Figure 1(a)). In constructing the spin Hamiltonian, we included both interaction groups, taking into account all neighbour interactions and their evolution with distance. As one can observe in Figure 1(b), both systems display FM characteristics, with Fe$_3$GaTe$_2$ exhibiting lower values compared to Fe$_3$GeTe$_2$ when considering solely the inter-plane couplings. However, more significant differences appear upon examination of in-plane exchanges. In Fe$_3$GeTe$_2$, the J$_{11}$ and J$_{33}$ interactions are antiferromagnetic (AF), resulting in a geometrically frustrated spin lattice [39,40]. Conversely, for Fe$_3$GaTe$_2$, J$_{11}$ and J$_{33}$ display values of 1.1 and -0.04 meV, respectively, indicating that the AF contribution of the in-plane couplings is almost suppressed, which in turn amplifies the net ferromagnetism. The values of T$_c$ derived from the calculated exchange couplings and MAE are 644 K for Fe$_3$GaTe$_2$ and 135 K for Fe$_3$GeTe$_2$. While our results predict an overestimated T$_c$ for Fe$_3$GaTe$_2$, they accurately reflect the comparatively higher critical temperature of Fe$_3$GaTe$_2$ over Fe$_3$GeTe$_2$.

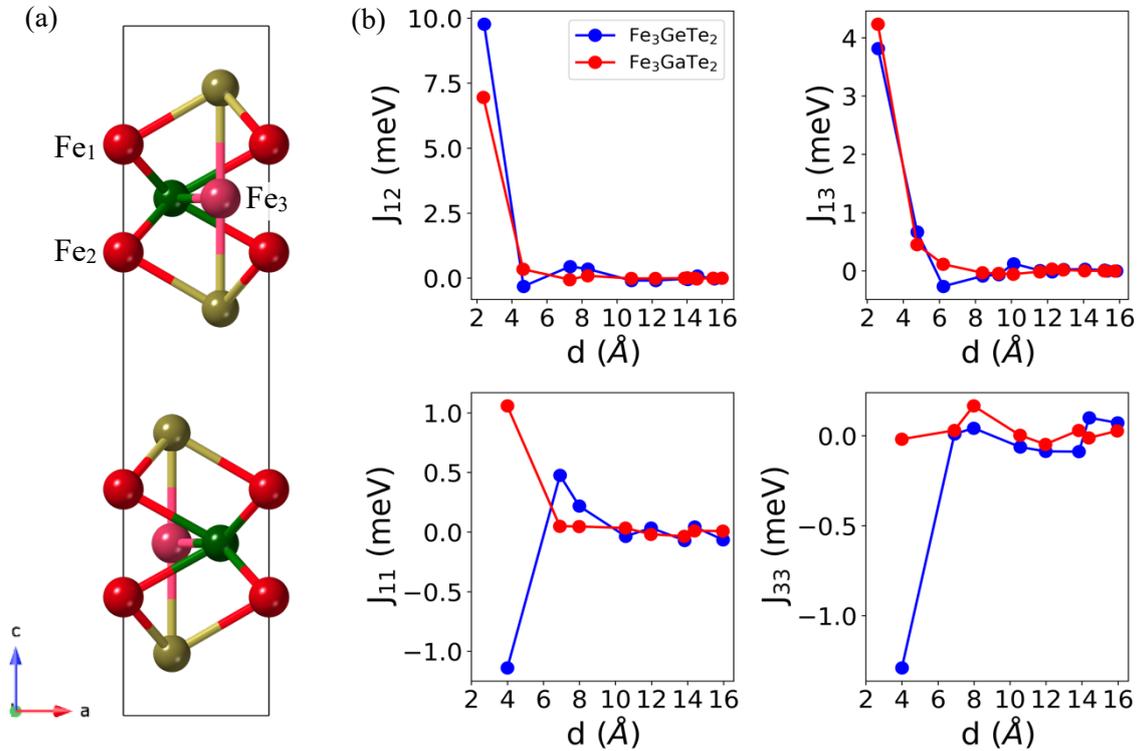

**Figure 1**. (a) Lateral view of a unit cell of bulk Fe$_3$GaTe$_2$, which includes two single-layers. Colour code: Fe$_{1,2}$ (red), Fe$_3$ (pink), Ga (green) and Te (yellow). (b) Inter-plane exchange interactions J$_{12}$, J$_{13}$ (top panel) and in-plane couplings J$_{11}$ and J$_{33}$ (bottom panel) for bulk Fe$_3$GaTe$_2$ (red) and Fe$_3$GeTe$_2$ (blue), as well as their evolution with distance to a maximum of 16 Å.

Considering the chemical and structural similarity between Fe$_3$GaTe$_2$ and Fe$_3$GeTe$_2$, which retains ferromagnetism and strong perpendicular anisotropy down to the monolayer limit [6,14], we investigate the electronic and magnetic properties of single-layer Fe$_3$GaTe$_2$. According to our first-principles calculations the optimized lattice parameters of the monolayer are a = b = 3.947 Å. These values are close to, but slightly smaller than, the abovementioned lattice constants of the bulk structure. We observe that single-layer maintains a FM ground state, with magnetic moments of 2.05 and 1.31 $\mu_B$/Fe for $Fe_{1,2}$ and $Fe_3$ atoms, respectively. This yields an average magnetic moment for Fe atoms of 1.80 $\mu_B$, which is nearly identical to the bulk and higher than the calculated for $Fe_3GeTe_2$ monolayer (1.61$\mu_B$).

Then, we explore the dynamic stability of $Fe_3GaTe_2$ monolayer by performing phonon calculations. As observed in Figure 2(a) the phonon dispersion presents no imaginary frequencies and thus the structure is stable, closely resembling the one of $Fe_3GeTe_2$ monolayer [41]. Furthermore, we study the energetic stability of $Fe_3GaTe_2$ monolayer calculating its formation energy relative to the bulk structure. This parameter is defined as the energy difference (per atom) between the monolayer and bulk forms, providing an estimation of the energy required to synthesize a single layer of material from its bulk counterpart. Using this definition, we obtain a value of 36.85 meV/atom, being close to the reported for $Fe_3GeTe_2$ monolayer (48.9 meV/atom) [41], and comparable to other vdW materials successfully exfoliated to the monolayer limit [42]. Regarding the electronic band structure and projected density of states (Figure 2(b)), one can observe that the system is metallic, with the $d$ orbitals of the transition metal playing an important role around the Fermi level. In contrast, the contributions Ga and Te are almost negligible, predominantly from the $p$ orbitals of these atoms.

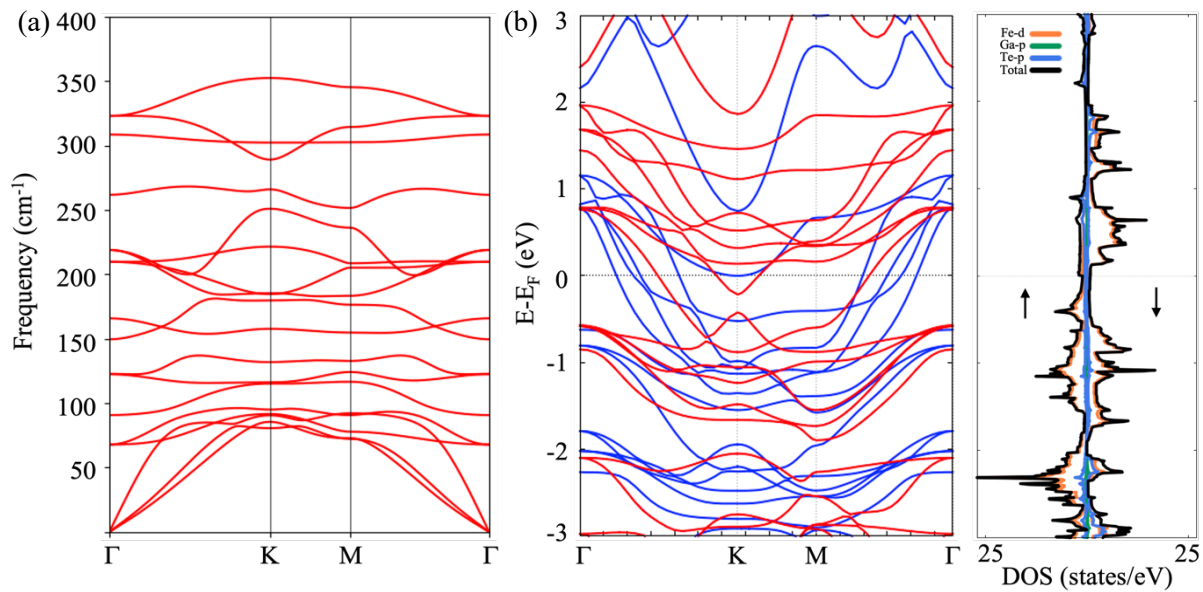

**Figure 2**. (a) Phonon spectrum, (b) electronic band structure (left) and orbital-resolved density of states (right) of $Fe_3GaTe_2$ monolayer. Blue (red) colour in the band structure indicates spin up (down) states.

Furthermore, we estimate the magnetic exchange interactions of $Fe_3GaTe_2$ monolayer. A direct comparison with the bulk results reveals minor variations, implying that the monolayer is likely to preserve a robust FM character (Figure S1). This observation is further supported by the calculation of

MAE, which yields a value of 0.36 meV/Fe, indicating a substantial perpendicular anisotropy with spin orientations aligned along the *c* axis. Our atomistic simulations predict a high $T_c$ value of 594 K, which is considerably higher than the one calculated for monolayer $Fe_3GeTe_2$ (100 K).

In Figure 3(a) we directly compare the exchange interactions in monolayers $Fe_3GaTe_2$ and $Fe_3GeTe_2$. Our analysis indicates that as in bulk, the in-plane couplings $J_{11}$ and $J_{33}$ are the origin of the observed differences in $T_c$ between these compounds. The distinct contributions originate from specific orbitals involved in the stabilization of long-range magnetic ordering in each system, as confirmed by our orbital-resolved analysis of the exchange parameters (Figure 3(b) and (c)). The ferromagnetic $J_{12}$ interaction is predominantly governed by $d_{z2}$-$d_{z2}$, $d_{xz}$-$d_{xz}$, and $d_{yz}$-$d_{yz}$ orbitals, with a small antiferromagnetic contribution arising from in-plane $d_{xy}$-$d_{xy}$ and $d_{x2-y2}$-$d_{x2-y2}$. The overall reduction in ferromagnetism within $J_{12}$ for $Fe_3GaTe_2$ can be attributed to a diminished FM contribution from $d_{xz}$-$d_{xz}$ and $d_{yz}$-$d_{yz}$ orbitals compared to the case of $Fe_3GeTe_2$ (Figure S2). This orbital-resolved description contrasts with the findings of Lee *et al*. [29], who associated the increase in $T_c$ in $Fe_3GaTe_2$ with a higher value of $J_{12}$ relative to $Fe_3GeTe_2$. Additionally, the FM nature of $J_{11}$ is attributed to a predominant FM $d_{yz}$-$d_{xz}$ superexchange pathway mediated by $p_y$ and $p_x$ orbitals of Ga and Te, respectively. This mechanism is highly diminished in $Fe_3GeTe_2$, eventually turning to become AF (Figure 3(b)). In addition, the $J_{33}$ in $Fe_3GeTe_2$ is primarily sourced from an AF $d_{z2}$-$p_x$-$d_{z2}$ mechanism mediated by Ge, which is nearly supressed in $Fe_3GaTe_2$ (Figure 3(c)). Our findings discard the possibility that structural differences between both systems are the determinant factor to explain the variations in magnetic exchange couplings and the differences in $T_c$ (Figure S3).

In order to assess the impact of $J_{11}$ and $J_{33}$ on the FM stabilization, we carry out calculations considering interactions until 16 Å (Figures S4, S5). In the case of only taking into account interactions up to 3 Å (thus only accounting for $J_{12}$ and $J_{13}$) we obtain values of $T_c$ of 251 K for $Fe_3GaTe_2$ and 220 K for $Fe_3GeTe_2$. This slight variation in $T_c$ between both systems suggests that a model restricted to nearest-neighbours inter-plane couplings is not adequate to describe the magnetic behaviour. Extending to 4 Å, where $J_{11}$ and $J_{33}$ are also considered, the $T_c$ of $Fe_3GaTe_2$ is enhanced up to 434 K, whereas the $T_c$ of $Fe_3GeTe_2$ is drastically reduced to 0 K. This proves the key role of in-plane couplings and the crucial significance of the AF character of those in the lower $T_c$ of $Fe_3GeTe_2$. Finally, expanding interactions to 16 Å provides values of $T_c$ of 594 and 100 K, highlighting the importance of long-range magnetic couplings.

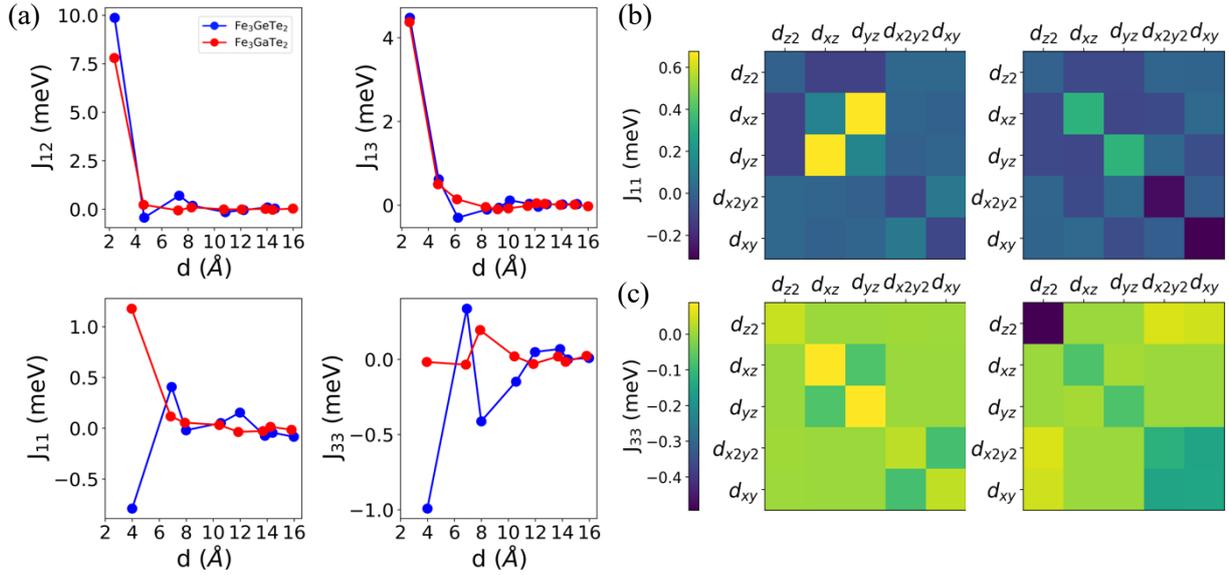

**Figure 3**. (a) Inter-plane exchange interactions $J_{12}$, $J_{13}$ (top panel) and in-plane couplings $J_{11}$ and $J_{13}$ (bottom panel) for monolayers $Fe_3GaTe_2$ (red) and $Fe_3GeTe_2$ (blue) along with their evolution with distance to a maximum of 16 Å. Orbital-resolved in-plane exchange parameters $J_{11}$ (b) and $J_{33}$ (c) for $Fe_3GaTe_2$ (left) and $Fe_3GeTe_2$ (right) monolayers.

Additionally, we investigate the evolution of the magnetic properties of $Fe_3GaTe_2$ monolayer upon mechanical deformation and electrostatic doping. Figures 4(a), (b) and (c) illustrate the variation of the magnetic moments for both Fe and ligands, along with the dependence of MAE and exchange parameters under biaxial strain. Notably, under strain (-4% to 4%), the average Fe magnetic moments steadily increase, while they remain almost null in the ligands. The MAE results reveal that the magnetization easy axis remains off-plane in the entire range of study. Its value is almost null at -4% and reaches a maximum at 1%, suffering a decrease at values of ε > 1%. According to our simulations, compression values slightly exceeding ε = -4% would induce in-plane magnetization within the system. On the other hand, a more substantial tensile strain would be required to modify the off-plane magnetization easy axis. Besides the evolution of MAE reflects the trend observed in the magnetic moments of $Fe_3$, the change in the latter is considerably smaller than the variation in MAE. This strongly suggests that the drastic change in MAE is not primarily due to spin moment variations. Prior studies have attributed this phenomenon with band shifts involving significant spin-orbit coupling [43]. This is further supported by our band structure calculations at various strain levels, demonstrating an effective modulation of the electronic properties around the Fermi level (Figure S6). In contrast, the evolution of the MAE diverges from that reported for $Fe_3GeTe_2$, where it parallels the evolution of magnetic moments and increases continuously upon tensile strain [41]. In addition, we can observe that magnetic couplings are highly sensitive to mechanical deformation. In Figure 4(c) we show that the FM exchange couplings exhibit an increase (decrease) upon compression (elongation) over the studied range. Notably, at $\varepsilon < -2\%$, $J_{12}$ significantly increases while $J_{11}$ decreases, the latter transitioning towards an AF state. We attribute these anomalous evolutions to the different trend of magnetic moments for $\varepsilon < -2\%$ compared to the rest of the map. Note that $Fe_1$ and $Fe_2$ atoms are situated within same *xy* plane and consequently the distance $Fe_1$-$Fe_2$ remains intact upon applied strain. However, the angle governing the $Fe_1$-Ga-$Fe_2$ superexchange pathway is reduced from 57.1º (-4%) to 53.3º (+4%) upon tensile strain, thereby having an impact in the coupling $J_{12}$ (see Figure S7 for further details). Based on the results of MAE and exchange parameters we computed the $T_c$ at different levels of strain (Figure 4(d)). For $\varepsilon = -4\%$ we observe that besides $J_{12}$, $J_{13}$ and $J_{33}$ are enhanced with respect the undistorted structure, the rapid decrease of $J_{11}$ results in a drop of the $T_c$ to a value of 320 K. Conversely, a 4% elongation yields a $T_c$ reduction by 10%.

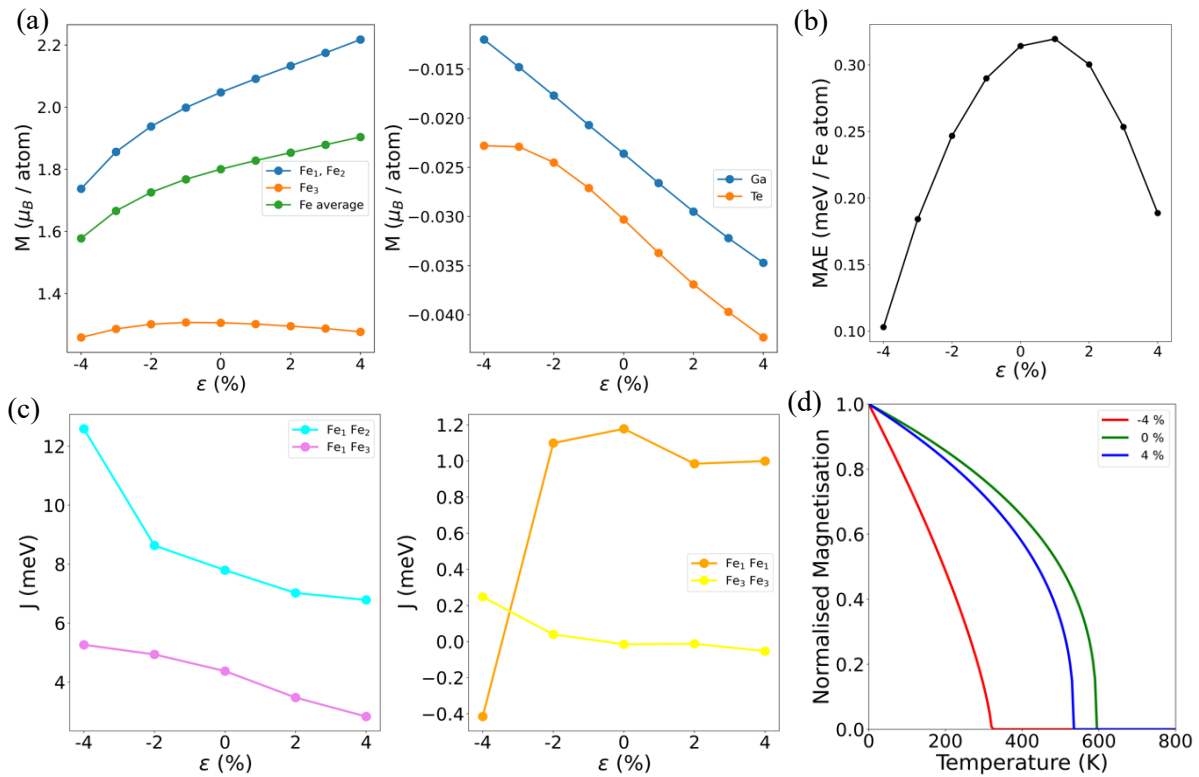

**Figure 4**. Evolution of (a) magnetic moments of metals and ligands (left and right panels, respectively), (b) MAE, (c) inter- (left) and in-plane (right) exchange parameters and (d) $T_c$ of $Fe_3GaTe_2$ monolayer upon applied strain.

The analysis of the impact of electrostatic doping on the magnetic properties shows that the magnetic moments of both equivalent and inequivalent Fe atoms are similarly affected, increasing and decreasing under electron and hole doping, respectively (Figure 5(a)). In contrast, the magnetic moments of Ga remain almost unchanged, whereas those of Te experience a pronounced increase under electron doping. Regarding the evolution of MAE reported in Figure 5(b), one can observe that the magnetization easy axis changes from off-plane (positive sign) to in-plane (negative sign) magnetization with a doping level around 0.15 h/f.u. This result agrees well with the theoretical observations of in-plane magnetism in bulk $Fe_3GaTe_2$ upon hole doping [29]. The evolution of magnetic exchange couplings is reported in Figure 5(c). We observe that electron doping leads to an increase of the FM interactions $J_{12}$ and $J_{13}$, suggesting a potential rise in $T_c$. However, there is a softening of the FM character of $J_{11}$ and $J_{22}$ that results in a reduced $T_c$ of 501 K (Figure 5(d)). This contrasts with the increase of the critical temperature reported for $Fe_3GeTe_2$ upon electron gating [14]. On the other hand, hole doping leads to a slight reduction in $J_{12}$ and $J_{13}$ compared to the undoped system, while $J_{11}$ and $J_{33}$ become more FM, resulting in an almost unchanged critical temperature. This aligns with recent experimental findings in bulk $Fe_{2.84}GaTe_2$, where Fe deficiency is linked to a slightly lower $T_c$ with respect to $Fe_3GaTe_2$ [33].

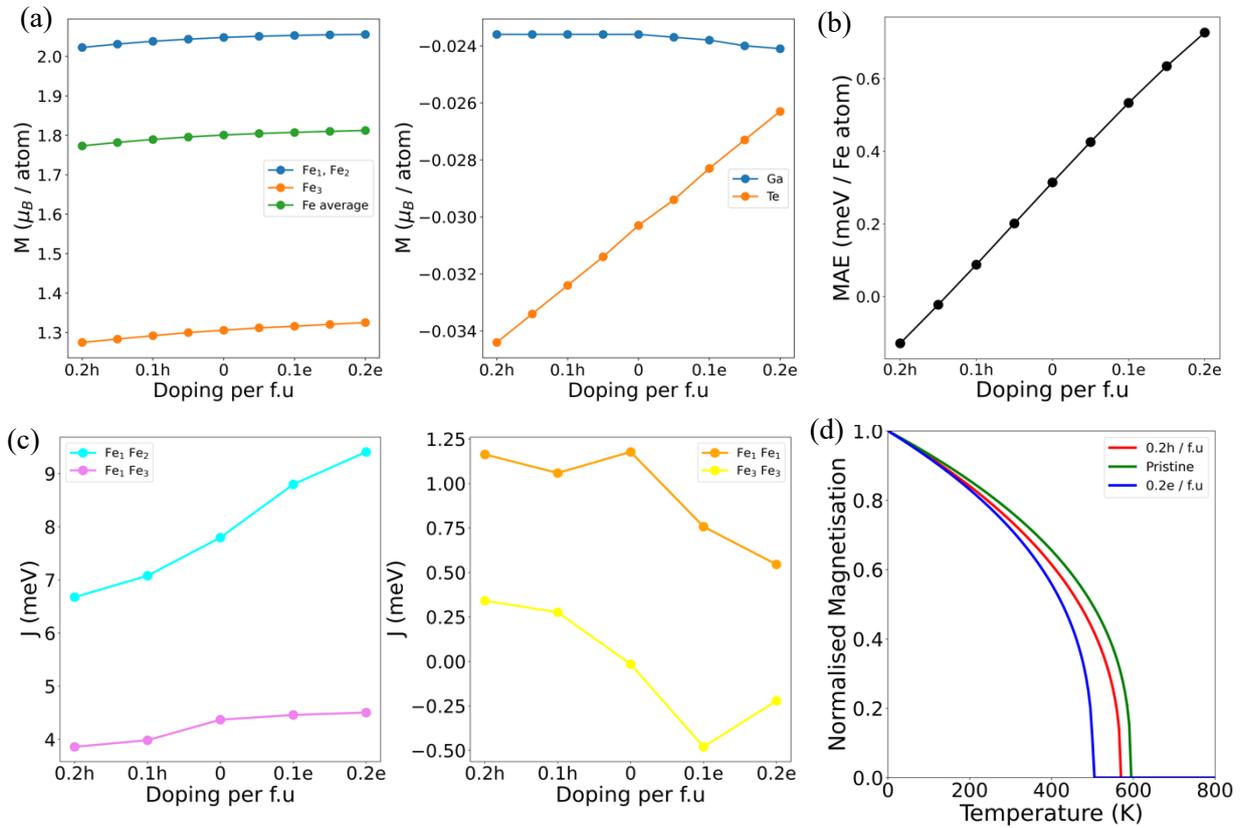

**Figure 5**. Evolution upon hole and electron doping of (a) magnetic moments of Fe, Ga and Te (left and right panels, respectively), (b) MAE, (c) inter- (left) and in-plane (right) exchange parameters and (d) $T_c$ of $Fe_3GaTe_2$ monolayer.

Finally, we examine the evolution of the electronic band structure with dopings of 0.2h/f.u. and 0.2e/f.u., observing that there is an effective shift of the bands around the high symmetry points Γ and K (Figure 6). These particular bands are postulated to significantly influence the MAE [29,43]. At zero doping, we note the presence of hole (around Γ) and electron (around K) pockets, which are mostly formed by *p* orbitals of Te and *d* orbitals of Fe atoms, respectively. Both pockets are shifted upwards (downwards) upon hole (electron) doping. Specifically, we note the presence of two electron pockets at K that contribute to an enhanced positive MAE. The first one (situated in the vicinity of K) suffers an effective downwards shift with increasing electron density, leading to heightened perpendicular anisotropy up to a value of 0.73 meV/Fe atom at 0.2e/f.u. In contrast, upon hole doping, the electron pockets rise in energy, with one band no longer intersecting the Fermi level. This results in a weakened perpendicular anisotropy and a preferential in-plane spin direction (-0.13 meV/Fe atom).

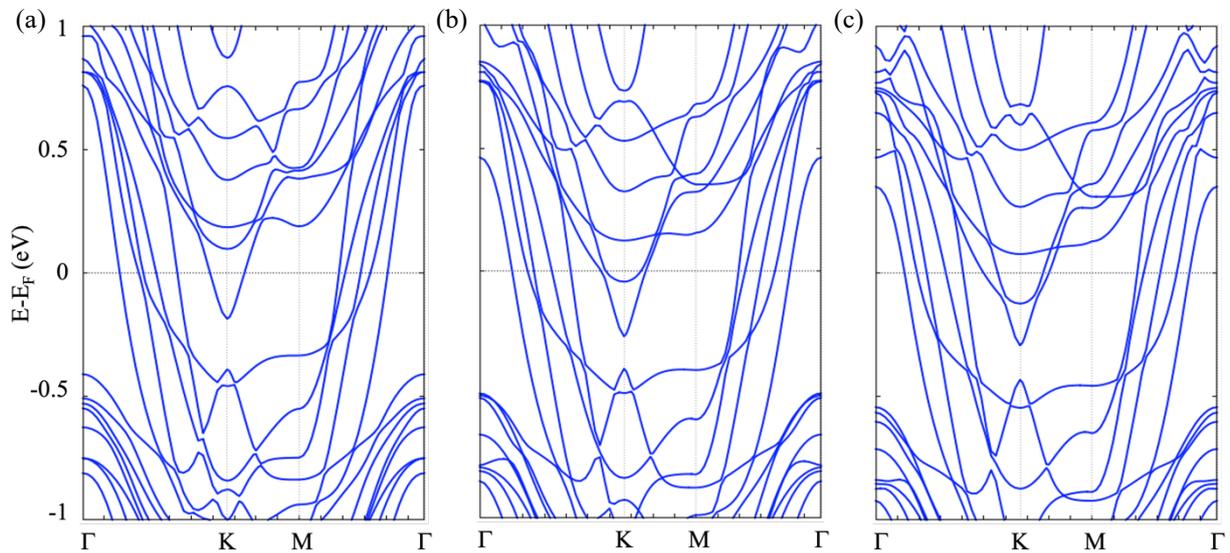

**Figure 6**. Electronic band structure for (a) 0.2h/f.u, (b) pristine and (c) 0.2e/f.u doped $Fe_3GaTe_2$ monolayer including spin-orbit coupling (SOC) effects.

**CONCLUSIONS**

In summary, we have investigated the magnetic and electronic properties of the above-room-temperature van der Waals ferromagnet $Fe_3GaTe_2$ via first principles calculations. By conducting a direct comparison between $Fe_3GaTe_2$ and $Fe_3GeTe_2$, we uncover the intricate microscopic mechanisms underlying their magnetic behaviour. Our analysis highlights the critical role of antiferromagnetic-ferromagnetic in-plane exchange interactions and their contribution to the higher $T_c$ observed in $Fe_3GaTe_2$ when contrasted with $Fe_3GeTe_2$. Moreover, our phonon calculations demonstrate the dynamical stability of single-layer $Fe_3GaTe_2$, while we predict strong perpendicular anisotropy and a high $T_c$, thus incentivising its isolation. Finally, we show that the exchange interactions of Fe$_3$GaTe$_2$ monolayer can be modified by strain engineering and electrostatic doping and prove their potential to tune the anisotropy of the system. Our findings lay a foundation for comprehending the origin of the critical temperature $T_c$ in both Fe$_3$GaTe$_2$ and Fe$_3$GeTe$_2$ and their future manipulation in magnetic and spintronic devices.

**COMPUTATIONAL DETAILS**

We carried out spin polarized density-functional theory (DFT) using the Quantum ESPRESSO package [44]. Spin-polarized local density approximation (LDA) [45] was used to approximate the exchange-correlation functional given that it has been proved to properly describe properties of Fe$_3$GaTe$_2$ and Fe$_3$GeTe$_2$ [14,36,46], avoiding an overestimation of magnetic moments [41]. For the bulk structures we relaxed the atomic coordinates while for the monolayers both atomic coordinates and lattice parameters were optimized. In both cases, the optimizations were carried out until forces on each atom were smaller than $1\cdot 10^{-3}$ Ry/au and the energy difference between two consecutive relaxation steps was less than $1\cdot 10^{-4}$ Ry. The electronic wave functions were expanded with well-converged kinetic energy cut-offs for the wave functions (charge density) of 75 (850) Ry. To properly describe the monolayers, a vacuum spacing of 18 Å was set along $c$ direction to avoid unphysical interactions between layers. For the bulk(monolayer) structures the Brillouin zone was sampled by a fine Γ-centered 10×10×3 (10×10×1) k-point Monkhorst–Pack, that was expanded to 15×15×3(15×15×1) for the calculations of MAE. The phonon spectrum was computed using a 3x3x1 supercell by means of the Phonopy code [47]. A tight-binding model based was constructed based on maximally localized Wannier function as implemented in the Wannier90 code [48]. Our reduced basis set is formed by the $d$ orbitals of Fe and $p$ orbitals of Ga, Ge and Te. Magnetic exchange interactions were determined using Green's function method as implemented in TB2J code [38] employing a 30×30×5 (30×30×1) supercell for the bulk (monolayer) structures. The T$_C$ was obtained by performing atomistic simulations as implemented in the VAMPIRE code [49].

To validate the obtained magnetic exchange couplings determined by the plane wave method, we double checked our results using a localized atomic orbital approach as implemented in the SIESTA package [50] (Figures S12 and S13), in which the local density approximation (LDA) was used to describe the exchange correlation energy [51,52]. We used a double-ζ basis set for all atoms and core electrons were described using norm-conserving Troullier−Martins pseudopotentials. A real-space mesh cutoff of 500 Ry and a 64x64x10(64x64x1) Monkhorst−Pack k-point mesh was used for the bulk (monolayer) calculations.

## ACKNOWLEDGMENTS

The authors acknowledge the financial support from the European Union (ERC-2021-StG101042680 2D-SMARTiES) and the Excellence Unit "María de Maeztu" CEX2019-000919-M). J.J.B acknowledges the Generalitat Valenciana (grant CIDEGENT/2023/1) and A.M.R. thanks the Spanish MIU (Grant No FPU21/04195). The calculations were performed on the Tirant III cluster of the Servei d'Informática of the University of València.